# Space Weather Effects on Critical Infrastructure

Gábor Facskó[1,2], Gergely Koban[2,3], Nikolett Biró [2,3], Munkhjargal Lkhagvadorj[2,4]

[1]Department of Informatics, Milton Friedman University, Budapest, Hungary.
[2]Department of Space Physics and Space Technology, Wigner RCP, Budapest, Hungary
[3]Doctoral School of Physics, Eötvös Loránd University, Budapest, Hungary.
[4]Faculty of Science, Eötvös Loránd University, Budapest, Hungary
`facsko.gabor@uni-milton.hu`

**Abstract.** Gas pipelines, transmission lines, overhead wires, transformers, GNSS navigation, and telecommunication systems are part of critical infrastructure. Industry, transportation, service operations, farming, and everyday life highly depend on this infrastructure. However, these systems are very sensitive to solar activity. Therefore, all activities above are vulnerable and defenseless against the catastrophic changes in Earth's cosmic environment. The Solar System is dominated by the influence of our star. In the Solar System, all objects are gravitationally bound and the radiation of the Sun provides the energy for example for the terrestrial biosphere. A small fraction of the energy produced in the core of our star turns into a magnetic field and emits the constant high-velocity flow, the solar wind. Solar magnetic activity produces radiation and ejects matter from the upper atmosphere of our star. The magnetic field of the solar wind interacts with the planetary magnetic fields and atmospheres.

These phenomena, called Space Weather have a serious influence on the radiation environment of Earth where telecommunication, Global Navigation Satellite System, meteorological, and other purpose satellites are located. The conductivity and transparency of the higher partly ionized atmospheric layer, the ionosphere also depend on solar radiation and activity. This fact makes the navigation and communication systems dependent on solar activity. Finally, the solar magnetic activity creates magnetic variations in the terrestrial magnetic field and induces currents in gas pipelines, transmission lines, overhead wires, and transformers. In this short briefing, we introduce the solar activity phenomena, and their influence on our planet's cosmic neighborhood and pro- vide a detailed description of the Space Weather effects on critical infrastructure. We describe the Hungarian national and global space weather forecast centres and capabilities. Finally, we share some guidelines on how to prepare for extreme space weather events.

## 1    Introduction

Here, the critical infrastructure (CI) means submarine internet cables, gas pipelines, transmission lines, overhead wires, transformers, Global Navigation Satellite System (GNSS), satellite telecommunication, and high frequency (HF) radio telecommunication systems. Submarine internet cables, gas pipelines, transmission lines, overhead wires, and transformers are long, conductive objects, therefore they are sensitive to quickly varied magnetic fields. GNSS navigation, satellite telecommunication, and HF radio telecommunication systems depend on the transmission and reflection ability of the ionosphere (that is a conductive layer of the upper atmosphere at 120 km altitude).

The Space Weather (SW) term has two meanings. It is a new(er) name for the research of the solar-terrestrial relationship. Additionally, the expression covers the efforts to predict the conditions of the near-Earth cosmic environment, the ionosphere, and the surface of our planet [1].

We have an aggressive and dominant neighbor: the Sun. The energy produced in the solar interior leaves our star by conductivity in its outmost layer. This conductive motion of the conductive solar plasma and the rotation of the Sun sets up and moves the solar dynamo that produces a strong magnetic field. The magnetic field of our star dominates the space nearby the Sun. This region of the space is called the heliosphere. The magnetic activity of the Sun makes serious disturbances in navigation and communication [1, 2].

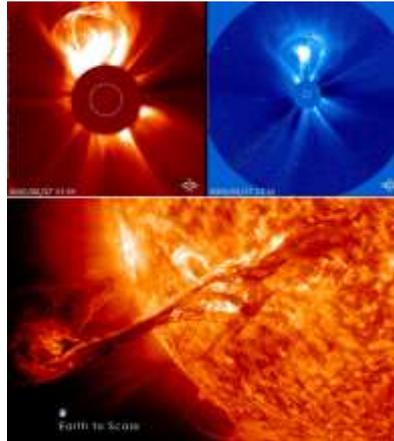

**Fig. 1.** (Top left and right) A coronal mass ejection on February 27, 2000, taken by SOlar and Heliospheric Observatory (SOHO) Large Angle and Spectrometric Coronagraph Experiment (LASCO) C2 and C3. A CME blasts into space a billion tons of particles traveling millions of miles an hour [32, 33]. (Bottom) Solar Dynamic Observatory (SDO) image of the Earth to scale with the filament eruption on August 31, 2012 [34]. (Credit: SOHO ESA & NASA; NASA Goddard Space Flight Center)

Sometimes two solar flux tubes with opposite directions situated close to each other gradually form an "X" shape configuration. If these flux tubes are close enough to each other and energetically better configuration arises, where the two halves of these opposite tubes form a shorter, bent structure. This phenomenon is called reconnection and it frees up a huge amount of energy. This region of the solar atmosphere reaches 15-20 million K. A bright flash in visible light, X-ray, and radio bursts are ejected. Charged particles are accelerated to high energy, these are the Solar Energetic Particles (SEPs).

A jet is ejected to the solar surface and usually (but not always) a huge amount of hot plasma is launched to the heliosphere. Its name is coronal mass ejection (CME) or solar storm (Fig. 1). The ejected plasma node remains connected magnetically to the Sun during its travel in the Solar System (Fig. 1, bottom). CMEs can also occur independently. A large CME could contain a billion tons of plasma that can be accelerated to several thousand km per second. Their size could be much larger than the diameter of the Sun. Therefore, solar material ejections fly through the interplanetary space, impacting any planet or spacecraft in their path [1, 2].

The reason for all troubles in the heliosphere is the atmosphere of the Sun. The temperature of the photosphere is 6000 K. The temperature of the chromosphere which is above the photosphere is around 10000 K. Finally, the temperature of the corona which exists above the chromosphere is around 1000000 K. You can see that in the solar atmosphere, the outer layers have higher temperatures. A strong, fast plasma flow originates from the outer layer of the solar corona, the solar wind. The distribution of the solar wind speed has two maxima at 400 km/s and 800 km/s [3]. These types of solar wind are called slow and fast solar wind or slow and fast solar wind streams.

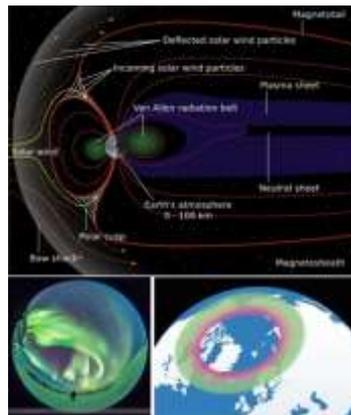

**Fig. 2.** (Top) The terrestrial magnetosphere and its regions: the bow shock, the magnetosheath, the magnetopause, the radiation belts, the plasmasphere, the cusps, the tail, and the neutral sheets. (Bottom left) The aurora looks beautiful on the night skies of high altitudes. The human eyes can usually observe the green light only. (Bottom right) The solar wind enters the terrestrial magnetosphere and creates aurora. There is aurora in daylight also, however, it cannot be seen. The aurora looks oval on the surface of the Earth around the north and south poles.
(Credit: Kiddle; Amazing Sky Photography; Discover the World 2023)

Our planet has a strong magnetic field that is tilted relative to the Ecliptic (the plane of Earth's orbit around the Sun) and bipole (that means it looks like the magnetic field of a bar magnet). The region where this magnetic field dominates is called the magnetosphere (Fig. 2, top). The solar wind flows faster than the sound speed and the speed of the perpendicular magnetic waves (the Alfvén speed). This flow grabs the magnetic field of the former solar plasma (the magnetic field is frozen into the solar wind plasma). This magnetic field cannot enter the magnetosphere. The information about the obstacle cannot propagate against the supersonic and super–Alfvénic flow because its maximal propagation speed is the sound and Alfvén–speed. Furthermore, the solar wind cannot hit the magnetosphere without any deceleration. Therefore, a shock, the so-called bow shock forms before the magnetosphere (Fig. 2, top). The solar wind passes slow down, its density and magnetic field magnitude increases, its temperature and entropy increase becomes more turbulent, and flows around the magnetosphere. Additionally, the solar wind direction changes and flows around the terrestrial magnetosphere. The deceleration region after the bow shock is called the magnetosheath (Fig. 2, top). The separation layer between the magnetosheath and the magnetosphere is called the magnetopause (Fig. 2, top). The night side of the magnetosphere (that is antisunward) is elongated and extended far beyond the lunar orbit. This region is called the tail (Fig. 2, top). The terrestrial tail is situated according to the direction of the solar wind as a windsock [1, 4, 5].

The solar wind enters the terrestrial magnetosphere and creates aurora and the radiation belts (Fig. 2, bottom left). The charged particles of the solar wind always enter the magnetosphere at the north and south magnetic poles of our planet. These regions are called cups (Fig. 2, top). These particles are trapped and rotate around the magnetic field; furthermore, the rotating high-energy particles also move along the magnetic field lines. When these particles approach a magnetic pole they bounce back. Therefore, the charged particles move between the magnetic poles. They also drift perpendicularly to this movement. If you add their location and speed you get a so-called ring current above the magnetic equator. The region of these charged particles is called radiation belts and Van Allen belts [1]. The ring current influences the magnetosphere of the Earth. The charged particles hit and excite the molecules and atoms of the atmosphere. These excited atoms eject visible (green) light (Fig. 2, bottom left). This light is called aurora [1]. The aurora activity is permanent under the aurora oval (Fig. 2, bottom right). The intensity of the aurora is proportional to the bandwidth of the radio communication through the ionosphere (Kjellmar Oksavik, personal communication).

On the boundary of Earth's upper, partially ionized, therefore conductive atmosphere (the ionosphere) and the lower magnetosphere, huge current systems are indicated [6]. These currents and the high-energy particles ejected by solar flares or accelerated CME shocks interact with the ionosphere and might change its transparency and reflectivity (Fig. 3). Therefore, solar activity influences GNSS navigation, satellite, and HF radio telecommunication systems [1]. Both observation and simulation of the variations of the terrestrial magnetic field in this region of our planet are challenging [7]. However, using ground-based magnetometers it is quite easy to record and monitor magnetic disturbances. Based on the measurements of geomagnetic observatories located at high and lower altitudes, you could determine whether a CME or another phenomenon disturbed and caused strong magnetic variations in the magnetosphere [1]. Based on these results the auroral activity could be predicted and currents induced by the rapid variations of the magnetic field could be calculated. These currents are called geomagnetically induced currents (GICs) and their magnitude depends on the conductivity of the surface materials [1]. However, the rapidly variable magnetic field also generates currents in all long conductive objects; such as gas pipelines, electric transmission lines, train over- head wires, submarine internet cables, and transformers [1]. Therefore, solar activity might cause damage to this type of critical infrastructure (Fig. 3).

## 2 Space Weather effects on critical infrastructure

The space weather phenomena might have serious effects on the critical infrastructure defined. The sources of the disturbances (Fig. 3, white ellipses) are the radiation outside of the Solar System (so-called cosmic rays), the SEPs, the radiation of solar flares, the solar flare radio bursts, the CMEs, and energetic particles from the radiation belts. These phenomena affect the ionosphere and induce currents in the objects on the surface of the Earth (Fig. 3, red rectangles).

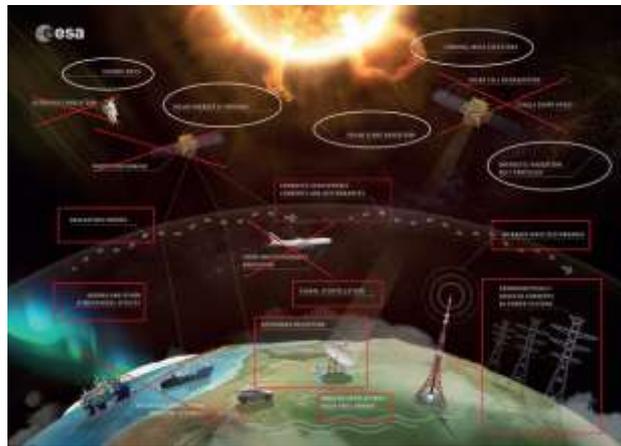

**Fig. 3.** (White ellipses) Space weather is guided by the radiation from outside the Solar System (cosmic rays), solar energetic particles, solar flare radiation (for example X–rays), radiation from the Earth's radiation belts, and the coronal mass ejections. (Red rectangles) These disturbances cause navigation errors, enhanced ionospheric currents and disturbances, high- frequency radio communication disturbances, GNSS location problems, radio reception disturbances, aurora, and, geomagnetically induced currents in current systems, gas pipelines, and train over-head wires. (Red crosses) These space weather effects are not interpreted in this paper. (Credit: ESA)

The ionosphere reflects high–frequency (HF) waves. Without this ability, HF radio broadcasts and communications would not be possible on Earth. This capability could be reduced and even destroyed by space weather phenomena. However, radio communication is also possible through this layer. Therefore, satellite communication and location determination using the GNSS position system are possible. The transparency of the ionosphere could be decreased by the space weather phenomena. Hence, satellite communication becomes problematic. Aurora is a beautiful tourist attraction and an important indicator of solar activity (Fig. 3, red rectangles). Furthermore, the intensity of the aurora is inversely proportional to the band- width of the radio communication through the ionosphere. The observation of the aurora is easy using optical full–sky cameras. The disturbed ionosphere could change the phase and the amplitude of the GNSS signals too. The name of this effect is scintillation which might cause errors in the navigation systems. Finally, the reception of radio waves could be disturbed (Fig. 3, red rectangles).

The rapidly variable terrestrial magnetic field induces geoelectric fields and currents on Earth's surface (Fig. 3, red rectangles). Furthermore, it creates geomagnetically induced currents in power systems, submarine cables, gas pipelines, and train overhead wires (Fig. 3, red rectangles). This is a dangerous effect. This phenomenon could block or destroy communication and energy transfer and the induced currents in a gas pipeline reduce the life expectancy of the pipes. These systems could be protected if we knew the magnitude of the effects or switched off if we could forecast the onset time of such an event.

## 3    A catastrophic Space Weather event

The Carrington Event was the most intense observed geomagnetic storm. Its maximum peaked from September 1 to 2, 1859. The geomagnetic storm was the result of a CME from the Sun hitting the magnetosphere of our planet [8]. The CME associated with a very bright solar flare on September 1, 1859, traveled 17.6 hours to the Earth [9]. The event causes strong aurora globally [10]. The aurora was visible from the poles to low latitude areas such as south-central Mexico, Queensland, Cuba, Hawaii, Japan, and China, and even very close to the equator in Columbia [11–15]. Because of the geomagnetically induced cur- rent from the rapidly changing terrestrial magnetic field, telegraph systems in Europe and North America failed, in some cases electrocuting their operators [16]. Some telegraph stations sparked and even caught fire [17]. Some operators could continue to send and receive messages after disconnecting their power supplies [18].

Such strong flares and CME ejections (or solar storms) occur quite rarely on our star. However, we can be sure that it will happen again. Nowadays, its impact would be catastrophic for our critical infrastructure. The geostationary and GNSS satellites would be damaged permanently. Therefore, communication and navigation would be impossible for a while. The transformer stations would be damaged and all computer systems, all elevators, all air conditioning, and all–electric equipment will be useless. The train, tram, and underground traffic would stop. The electric transfer systems would be useless. The gas pipelines would have to be shut down because of security reasons. Human civilization would be thrown back to the 19th century for months. The effects of such a solar storm (or CME) look similar to a global thermonuclear bombardment exploding at high altitudes inside the magnetosphere using charges with improved microwave radiation capabilities. The researchers at Lloyd's of London and Atmospheric and Environmental Research (AER) in the US estimated how much the restoration of damages would cost if Earth was hit by a Carrington event size CME. The damage would range from $600 billion to $2.6 trillion in the US alone, which was 3.6 to 15.5 percent of annual GDP in 2013 [19]. Therefore, the authorities must be prepared to recognize the effects of such a catastrophic event using national resources. Because these effects would be global, international communities could not rally and coordinate disaster response efforts such as in the case of classic, localized dis- asters (e.g. earthquakes), and every country would be left to deal with the damages alone.

## 4      Space Weather forecast centres

Sometimes the space weather effects have a catastrophic influence on critical infrastructure. For example, a solar storm observed by STEREO in July 2012 was a CME of comparable magnitude to the one causing the Carrington Event. However, smaller effects could be also highly devastating because our civilization is highly sophisticated and globalization connected the far regions of the world. Therefore, many nowcast and prediction centres exist on Earth. The aim of the forecast seems to be clear. It is also important to know the recent, and current conditions of the terrestrial cosmic environment. This dedicated task uses the same methods and detectors as the forecast systems. Hence, we call this monitoring activity nowcast. The most famous related organization is the National Aeronautics and Space Administration (NASA) Goddard Space Flight Center (GSFC) Space Weather Laboratory located in Green- belt, Maryland, USA. In GSFC staff develops forecast methods and does general space science research too. NASA launched and is operating the Advanced Composition Explorer (ACE) and the Deep Space Climate Observatory (DSCOVR) solar wind monitoring satellites [20, 21]. In the USA not only NASA maintains a space weather prediction centre. The Space Weather Prediction Center National Oceanic and Atmospheric Administration (NOAA) is located in Boulder, Colorado. The Cooperative Institute for Research In Environmental Sciences (CIRES) and the University of Colorado also develop tools for a space weather forecast. In the United Kingdom the UK MET Office in Exeter, Devon provides space weather forecasts. The UK Met Office (as well as most other organizations) uses US spacecraft (ACE and DSCOVR) to get continuous (operational) solar wind data. One of the most important tools for modeling the heliosphere in 3D is the ENLIL [22], which was also developed in the USA. The ENLIL considers the solar wind as a magnetized fluid and predicts its behavior and flow inside and outside of Earth's orbit. This approach of the solar wind is called magnetohydrodynamic (MHD) description [23]. Recently the EUropean Heliospheric FORecasting Information Asset (EUHFORIA) MHD model was developed at the KU Leuven Centre for Mathematical Plasma Astrophysics, therefore Europe could be independent of the USA [24]. China also established its forecast organization in Beijing, State Key Laboratory of Space Weather (SKSW). In Brazil, the Instituto Nacional de Pesquisas Espaciais, Estudo e Monitoramento Brasileiro do Clima Espacial (INPE/EMBRACE) provides forecasts. The INPE develops and builds the Galileo Solar Space Telescope (GSST) to boost its solar and space weather activity forecast capabilities [25]. The South African National Space Agency (SANSA) has its space weather monitoring and forecast centre in Hermanus, Western Cape because the rapidly developing space industry of the country needs up-to-date predictions. Naturally, the European Space Agency (ESA) has the Space Safety Programme and its Space Weather Service Network (https://swe.ssa.esa.int). The headquarters of the program is located in Darmstadt, Germany but its operation centre is situated in Brussels. Various European and other countries founded the European Consortium for Aviation Space weather User Services (PECASUS). The PECASUS aims to improve the security of civil aviation.

## 5  Hungarian Space Weather capabilities

Based on the long list above, it seems not only great powers have the capability and sources to establish such a service. There- fore, it would be wise to establish the Hungarian National Space Weather Prediction Centre. Hungary has the necessary knowledge. In the Wigner Research Centre for Physics at the Department of Space Physics and Space Technology the main research fields are the heliosphere, comets, planetary magnetospheres, dusty plasmas, and MHD modeling. In the Centre for Energy Research Radiation Protection Department Space Dosimetry Research Group the engineers have long experience to build dosimeters (Pille, [26]), Langmuir Probes (plasma density and temperature instruments) constructing and recently they started developing magnetometers in collaboration with the Imperial College, London [27]. The Institute of Earth Physics and Space Science (EPSS) provides ionosonde, GIC, and geoelectric field observations in addition to possessing a huge database of solar spot observations). Based on this database a very effective flare forecast method was developed supported by the ESA [28, 29]. They also have experts on terrestrial bow shock and ground-based magnetometers. In the Eötvös Loránd University, the Space Research Group has experience in building electrostatic wave instruments and experts of the plasmasphere and the radiation belts [30, 31]. The Department of Astronomy is located on the same corridor at the same university, where some solar physicists work. Finally, at the Budapest University of Technology (BME) and Economics in the Department of Electron Devices and the Department of Broadband Infocommunications and Electromagnetic Theory the BME Space Technology Research Group develops pico- and pocket satellites and power supplies for space missions. This group built the MASAT (MO–72), the first Hungarian picosatellite, and the SMOG-1 and SMOG-P PocketQubes (1PQ). This list of capabilities and expertise is not perfect because some researchers and engineers work in various groups at the same time therefore their expertise cannot be coupled to a research group.

According to the list above the Hungarian space scientists and engineers possess the necessary knowledge for establishing a forecast and nowcast centre to predict conditions from Earth's surface to the heliosphere. Some of the researchers and engineers of these groups have already participated in the work of the ESA's Space Safety Programme. Therefore, founding a national space forecast centre is based on political will and funding only.

## 6  Summary

In this brief paper, we defined the terms of space weather and critical infrastructure, described the flares, coronal mass ejections (or solar storms), the solar wind, the magnetosphere (and its regions), the ionosphere, the geomagnetically induced currents, and the aurora. We provided information about the damage to critical infrastructure caused by space weather effects. We also described the most devastating solar storm in human history, the Carrington event, and interpreted the impact of a similar event on our technical civilization. We shared a list of international space weather prediction centres and emphasized the Hungarian space weather experts and capabilities. We suggest establishing an operational national space weather prediction centre. We urge the authorities to construct a national emergency scenario for extreme space weather effects for the police, defense forces, fire departments, and civil defense organizations. Their staff must be briefed on how to realize the extreme space weather impact and trained on how to restart the technical equipment after such destruction caused by our star.

**Acknowledgments.** This work was partially financed by the National Research, Development, and Innovation Office (NKFIH) FK128548 grant. ML was supported by the Stipendium Hungaricum Scholarship.